\documentclass[superscriptaddress,
 amsmath,amssymb,
 aps,
prl,
reprint
]{revtex4-2}

\usepackage{graphicx}% Include figure files
\usepackage{dcolumn}% Align table columns on decimal point
\usepackage{bm}% bold math
\usepackage{physics}
\usepackage{hyperref}
\usepackage{amsfonts, amsmath}
\usepackage{dsfont}
\usepackage{amsthm}
\usepackage{bm}
\usepackage{bbold}
\usepackage{color}
\usepackage{lipsum,babel}
\usepackage[normalem]{ulem}

\newcommand{\XE}[0]{{\text{XE}}}

\newcommand{\per}[0]{\text{Per}}

\begin{document}
\definecolor{navy}{RGB}{46,72,102}
\definecolor{pink}{RGB}{219,48,122}
\definecolor{grey}{RGB}{184,184,184}
\definecolor{yellow}{RGB}{255,192,0}
\definecolor{grey1}{RGB}{217,217,217}
\definecolor{grey2}{RGB}{166,166,166}
\definecolor{grey3}{RGB}{89,89,89}
\definecolor{red}{RGB}{255,0,0}

\preprint{APS/123-QED}

\title{Spoofing cross entropy measure in boson sampling}
\author{Changhun Oh}
%\email{changhun@uchicago.edu}
\affiliation{Pritzker School of Molecular Engineering, University of Chicago, Chicago, Illinois 60637, USA}
\author{Liang Jiang}
\affiliation{Pritzker School of Molecular Engineering, University of Chicago, Chicago, Illinois 60637, USA}
\author{Bill Fefferman}
\affiliation{Department of Computer Science, University of Chicago, Chicago, Illinois 60637, USA}

\begin{abstract}
Cross entropy (XE) measure is a widely used benchmarking to demonstrate quantum computational advantage from sampling problems, such as random circuit sampling using superconducting qubits and boson sampling (BS).
We present a heuristic classical algorithm that attains a better XE than the current BS experiments in a verifiable regime and is likely to attain a better XE score than the near-future BS experiments in a reasonable running time.
The key idea behind the algorithm is that there exist distributions that correlate with the ideal BS probability distribution and that can be efficiently computed. 
The correlation and the computability of the distribution enable us to post-select heavy outcomes of the ideal probability distribution without computing the ideal probability, which essentially leads to a large XE.
Our method scores a better XE than the recent Gaussian BS experiments when implemented at intermediate, verifiable system sizes. 
Much like current state-of-the-art experiments, we cannot verify that our spoofer works for quantum advantage size systems. 
However, we demonstrate that our approach works for much larger system sizes in fermion sampling, where we can efficiently compute output probabilities.
Finally, we provide analytic evidence that the classical algorithm is likely to spoof noisy BS efficiently.
\end{abstract}

%We study the XE benchmarking for BS.
%First, we derive the XE for ideal and noisy Fock-state BS and show how noise decreases the XE.

\maketitle

Quantum computers are believed to efficiently solve problems that classical counterparts cannot, such as integer factoring \cite{shor1994algorithms} and quantum simulation \cite{lloyd1996universal}.
Whereas scalability and fault tolerance are required to solve hard and practical problems, currently available devices are noisy intermediate-scale quantum (NISQ).
Hence, huge attention has been paid to demonstrating quantum advantage by exploiting NISQ devices.
Sampling problems are particularly promising for the demonstration thanks to rigorous evidence that classical computers cannot efficiently solve them \cite{aaronson2011computational, hamilton2017gaussian, bouland2019complexity, deshpande2021quantum}.
Indeed, we have recently seen the first claims of quantum advantage using random circuit sampling (RCS) with superconducting qubits~\cite{arute2019quantum, wu2021strong, morvan2023phase} and Gaussian boson sampling (GBS)~\cite{zhong2020quantum, zhong2021phase, madsen2022quantum, deng2023gaussian}.
However, the apparent limitation of current experiments is that they are not scalable because uncorrected noise decays quantum signal as the system size grows \cite{aharonov1996limitations, kalai2014gaussian, bremner2017achieving, gao2018efficient, renema2018classical, shchesnovich2019noise, garcia2019simulating, noh2020efficient, takahashi2021classically, qi2020regimes, oh2021classical, villalonga2021efficient, aharonov2022polynomial, oh2023classical, liu2023complexity, Oh2023Tensor}.
Thus, finding appropriate size experiments is crucial in which the system size is sufficiently large so that classical computers cannot efficiently simulate them but not too large for noise to annihilate quantum signals.

Finding an appropriate regime is also important to enable classical verification because verification techniques such as cross-entropy benchmarking (XEB) cost exponential time \cite{boixo2018characterizing}.
The state-of-the-art method is XEB, which is sample-efficient for RCS and GBS~\footnote{XEB is sample-efficient in the sense that it does not require an exponentially large number of samples while it is computationally inefficient because computing the ideal output probability requires exponential time.}; scoring a large XE is considered evidence of quantum advantage.
The premise behind the benchmarking is that appropriate-size experiments maintain sufficient quantum signals so that the experimental score cannot be attained by classical devices within reasonable costs due to the remaining quantum signals.
For RCS \cite{arute2019quantum, wu2021strong}, there have been extensive studies to support the premise \cite{boixo2018characterizing, arute2019quantum, aaronson2016complexity, aaronson2019classical}.
While we have some limited understanding of what XE measures in RCS \cite{gao2021limitations, boixo2018characterizing, arute2019quantum, aaronson2016complexity, aaronson2019classical, pan2021solving} with many interesting debates of XEB \cite{pan2021solving, gao2021limitations}, 
our understanding of XE in BS is much more limited than RCS.
Crucially, there is no theoretical evidence that attaining a high XE is classically hard, to the best of our knowledge.
Despite this, the XEB has been used in many state-of-the-art quantum advantage experiments~\cite{zhong2020quantum, zhong2021phase, madsen2022quantum, deng2023gaussian}. XEB is particularly attractive since the benchmark does not depend on an adversarial mock-up distribution. This {\it adversary independence} feature allows us to use XEB to benchmark noisy experiments in situations in which we are ignorant of the best possible classical spoofing algorithm, as is the case with GBS experiments~\footnote{Other benchmarking methods employed~\cite{zhong2020quantum, zhong2021phase, madsen2022quantum, deng2023gaussian}, such as Bayesian test and correlator methods, also do not have rigorous evidence for hardness. Also, the Bayesian test is genuinely implemented against adversarial samplers; thus, it cannot rule out all possible classical algorithms. We emphasize that the current GBS experiments employ XEB against certain adversarial samplers because we do not know the ideal score.}.
%it may be a promising benchmark that does not require an adversarial sampler to compare, which is crucial to rule out all possible efficient classical algorithms~\footnote{Other benchmarking methods employed~\cite{zhong2020quantum, zhong2021phase, madsen2022quantum, deng2023gaussian}, such as Bayesian test and correlator methods, also do not have rigorous evidence for hardness. Also, the Bayesian test is genuinely implemented against adversarial samplers; thus, it cannot rule out all possible classical algorithms. We also emphasize that the current GBS experiments employ XEB against certain adversarial samplers because we do not know the ideal score.}.}

In this Letter, we provide a heuristic classical algorithm that scores better than the current intermediate-scale GBS (in a verifiable regime) and is likely to score better than the near-future BS experiments for XEB.
First, we numerically demonstrate using a small-scale GBS that our algorithm selectively generates heavy outcomes of BS and scores better than the ideal distribution.
For larger systems in a quantum advantage regime, due to the inefficiency of estimating XE and a large computational cost, the frequently used method is to analyze an intermediate-size experiment in a classically verifiable regime, instead of the largest experiment.
Following this, we demonstrate that the XE of the proposed sampler achieves a significantly larger XE score than the intermediate-size GBS of the most recent experiments in Refs.~\cite{zhong2021phase, madsen2022quantum}.
To predict its behavior for large system sizes, we analyze its performance for fermion sampling~(FS), which is efficiently simulable and verifiable, and provide analytical evidence of efficient spoofing of noisy BS.
Therefore, our classical algorithm is expected to score better than the near-future BS experiments in a reasonable time.
We finally discuss other existing spoofing methods and benchmarking.

\begin{figure}[t]
\includegraphics[width=180px]{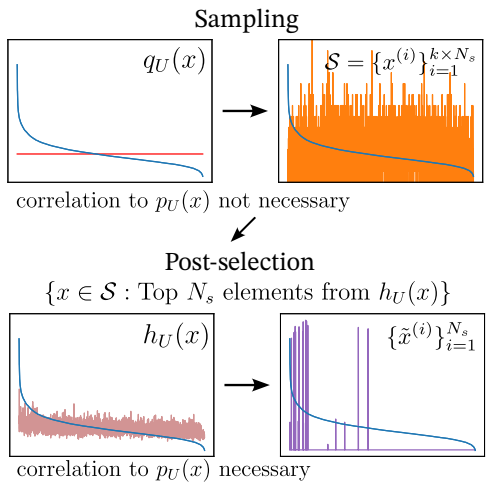}
\caption{Proposed algorithm for spoofing XEB. The blue curve represents the ideal distribution $p_U(x)$. See the main text for the detailed procedures.
}
\label{fig:scheme}
\end{figure}

\emph{XE.---}\label{sec:XEB}
The (log) XE of $q_U(x)$ with respect to the ideal probability $p_U(x)$ is defined and estimated as~\cite{boixo2018characterizing, arute2019quantum}
\begin{align}
    \text{XE}\equiv\sum_x q_U(x)\log p_U(x)\approx \frac{1}{N_s}\sum_{i=1}^{N_s}\log p_U(x^{(i)}),
\end{align}
where $N_s$ is the number of samples from $q_U(x)$, $q_U(x)$ is an experimental probability distribution or a mock-up probability distribution, and the sum is taken over all measurement outcomes $x$.
Since there are exponentially many different outcomes for BS, the XE of an experimental or a mock-up distribution with respect to an ideal probability distribution cannot be efficiently computed; thus it is estimated by sampling $\{x^{(i)}\}_{i=1}^{N_s}$ from $q_U(x)$ in practice (even estimation is inefficient because the cost for computing probabilities is exponential in the system size.).
Unlike RCS, due to the lack of understanding of the ideal score of BS, the XE was used as a relative quantity against mock-up distributions~\cite{madsen2022quantum}.
Specifically, if an experimental XE is larger than the mock-up distribution's, it implies that the former generates heavier outcomes, considered as evidence of quantum advantage for BS.
Such a method was used as evidence of quantum computational advantage in the recent GBS experiment \cite{madsen2022quantum}.

%\emph{Spoofing cross entropy benchmarking of Gaussian boson sampling.---}\label{sec:spoof}
\emph{Heavy outcome generation.---}\label{sec:spoof}
Now, we present a classical algorithm attaining a large XE.
For BS, there are probability distributions that can be efficiently sampled and correlate with the ideal probability distribution.
For example, fully distinguishable BS has nonzero correlation \cite{aaronson2013bosonsampling}.~Another example is a probability distribution having the same low-order marginal distributions, which was used to spoof a GBS experiment~\cite{villalonga2021efficient}.
Nevertheless, the correlations are too small to obtain a larger XE than the experiments \cite{madsen2022quantum}.
We now provide how to increase the XE from the small correlation.
Note that our scheme is not limited to BS in principle.

\iffalse
Let us explain how the algorithm works step by step, which is illustrated in Fig~\ref{fig:scheme}.
Algorithm 1:
{\bf Step~1.} Find a classical mock-up sampler $q_U(x)$ that is {\it efficiently samplable} and {\it correlates} with the ideal probability $p_U(x)$.
{\bf Step~2.} Generate $k\times N_s$ samples $\{x^{(i)}\}_{i=1}^{k\times N_s}$ from $q_U(x)$, where $N_s$ is the target number of samples and $k>1$ is a positive integer, which represents the post-selection rate.
The procedure is efficient by definition of $q_U(x)$.
{\bf Step~3.} Compute the probability of each sample with respect to the probability distribution $q_U(x)$.
In this procedure, we require the probability distribution $q_U(x)$ to be {\it efficiently computable}.
{\bf Step~4.} Post-select $N_s$ samples that have the $N_s$ largest probabilities $q_U(x)$ out of $k\times N_s$ samples.
{\bf Step~5.} Output the post-selected samples.
\fi

Let us denote as $q_U(x)$ a probability distribution for sampling and as $h_U(x)$ a distribution for post-selection.
%We emphasize that $h_U(x)$ does not need to be a probability distribution.
We now present the algorithm, illustrated in Fig.~\ref{fig:scheme}.
{\bf Step~1.} Choose an efficient classical sampler $q_U(x)$.
{\bf Step~2.} Generate $k\times N_s$ samples from $q_U(x)$.
{\bf Step~3.} Compute $h_U(x)$ for each sample.
In this procedure, we require $h_U(x)$ {\it efficiently computable} and {\it correlated} to $p_U(x)$.
{\bf Step~4.} Post-select and output $N_s$ samples whose $h_U(x)$'s are largest out of $k\times N_s$ samples~\footnote{Another possible way is to generate $k$ samples and post-select a single sample out of $k$ samples and iterate $N_s$ times, instead of generating $k\times N_s$ and post-select $N_s$ samples at once. While we do not expect a significant difference, we employ the latter in this work consistently.}.
%{\bf Step~5.} Output the post-selected samples.

The principle behind the post-selection is that due to the correlation between $p_U(x)$ and $h_U(x)$, we obtain samples likely to have a larger probability with respect to the ideal distribution $p_U(x)$ by selecting the samples with larger $h_U(x)$.
Thus, the most crucial step is to find $h_U(x)$ correlated with the ideal distribution but easy to compute.
Together with our numerical result below, the existence of such indicators might be related to hardness of heavy outcome generation in that the indicators enable us to generate heavy outcomes, where the heaviness means that its probability is larger than the median of probabilities \footnote{We note that generating outcomes whose probabilities are larger than the median is sufficient to spoof the XEB because even running a quantum device can only generate heavy outcomes in this sense slightly better than the light outcomes \cite{aaronson2019classical}.}.
Note that we do {\it not} compute the ideal probability in the entire procedure for sampling.
We provide more discussions about the choice of $q_U(x)$ and $h_U(x)$ in Ref.~\cite{supple}.
%Note that such a correlation indicator was first introduced in Ref.~\cite{aaronson2013bosonsampling} for a different purpose which is to distinguish a BS from uniform sampler.

\begin{figure}[t]
\includegraphics[width=230px]{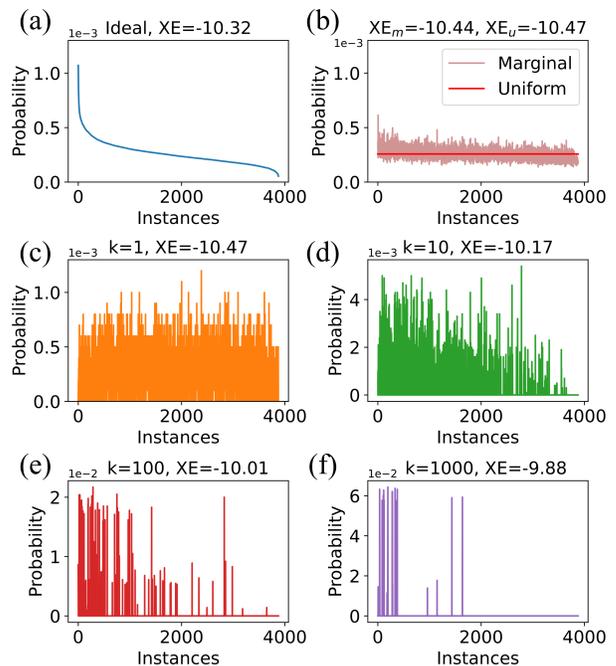}
\caption{(a) Ideal distribution $p_U(x)$ and (b) the distribution $h_U(x)$ are obtained by computing all probabilities of $N$ photon outcomes,
where $h_U(x)$ and $q_U(x)$ are sorted in the same order of $p_U(x)$.
%is also obtained by directly computing all probabilities.
Here, $\text{XE}_m$ and $\text{XE}_u$ represent the XE of $h_U(x)$ and $q_U(x)$, respectively.
(c)-(f) The spoofing with different $k$ with $N_s=10^4$.
%For $k=1$, we do not post-select, so that it is an empirical distribution from the uniform distribution.
%The number of samples for (c)-(f) is 
}
\label{fig:xanadu3}
\end{figure}

\emph{Spoofing XEB in GBS experiments.---}\label{sec:result}
%\subsection{Small system size}
To illustrate how the spoofing procedure operates in practice, we analyze the method for a small-size GBS circuit with the number of modes ${M=16}$ and the number of photons ${N=4}$~\cite{madsen2022quantum}.
Here, we choose uniform distribution $q_U(x)$ and first-order-marginal-based distribution $h_U(x)\equiv \prod_{i=1}^M p_U(x_i)$,
where $p_U(x_i)$ is the ideal marginal probability of the $i$th mode.
Here, the marginal probabilities can be easily computed since the reduced state is a single-mode Gaussian state whose covariance matrix is a submatrix of the full covariance matrix \cite{serafini2017quantum}.
Because the chosen $h_U(x)$ perfectly recovers the first-order marginals of the ideal distribution $p_U(x)$ by definition, it correlates with the ideal distribution.
%Since the number of possible $N$-photon outcomes $\binom{M+N-1}{N}=3876$ and the number of photons are small, 
After computing all the ideal probabilities $p_U(x)$, we sort the outcomes in descending order.
%Due to the small scale, we can compute all the probabilities.
We also compute $q_U(x)$ and $h_U(x)$ and sort them in the same order as the ideal case.
Although we use log XE for comparison with experiments in Refs.~\cite{zhong2021phase, madsen2022quantum}, a similar result using linear XE is provided in Ref.~\cite{supple}.
%For simulation, we use the number of samples as $N_s=10^4$ for all cases with different $k=1,10,100,1000$.
%Especially $k=1$ means that we do not post-select, i.e., the resultant distribution is the empirical version of the uniform distribution $q_U(x)$.
%For larger $k>1$, as explained above, we generate $k\times N_s$ number of samples from the uniform sampler and post-select the samples that have the $N_s$ largest values from the correlation indicator $h_U(x)$.

As shown in Fig.~\ref{fig:xanadu3}(a) and (b), the ideal probability $p_U(x)$ is likely to be large if $h_U(x)$ is large, which clearly shows their correlation, while each XE of $q_U(x)$ and $h_U(x)$ is smaller than the ideal distribution.
However, as the post-selection rate $k$ increases, the samples are concentrated on large probabilities of the ideal distribution, which clearly shows that our procedure selectively generates heavy outcomes.~Especially for $k=10^2, 10^3$, the samples are highly concentrated on the heavy outcomes.
Consequently, the spoofer's XE is larger than the ideal probability.
Therefore, one can attain a large XE by sampling heavy outcomes without directly simulating the desired circuit.
Here, the XE might not monotonically increase as the post-selection rate because one might end up selecting less heavy outcomes.
We also analyze the same procedure for Fock-state BS (FBS) using different $h_U$ in Ref~\cite{supple}.

\begin{figure}[t]
\includegraphics[width=240px]{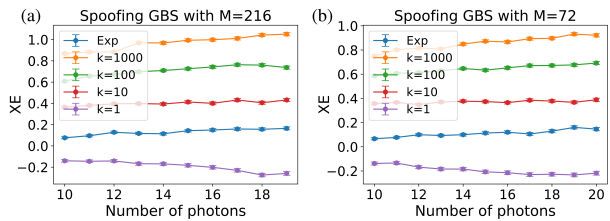}
\caption{(a) 216 modes with 216 squeezed states with $r\approx 0.89$. (b) 72 modes with 72 squeezed states with $r \approx 0.53$.
The error bar represents the standard deviation of $\log p_U(x)$ of obtained samples divided by $\sqrt{N_s}$, and $N_s=10^3$.}
\label{fig:xanadu}
\end{figure}

%\subsection{Intermediate system size}
Now, we consider intermediate-size circuits used for demonstrating quantum advantage in Ref.~\cite{madsen2022quantum}, where the experimental samples attain a larger XE than various mock-up distributions.
Since XE is not an efficient verification method, the result based on the intermediate-scale circuits is claimed to be evidence of the quantum advantage of their largest circuit.
Following Ref.~\cite{madsen2022quantum}, we normalize as $\text{XE}=\sum_{i=1}^{N_s} \log [p_U(x^{(i)})/\mathcal{N}]/N_s$,
%\begin{align}\label{eq:XE_norm}
%    \text{XE}=\frac{1}{N_s}\sum_{i=1}^{N_s} \log \frac{p_U(x^{(i)})}{\mathcal{N}},
%\end{align}
where $\mathcal{N}\equiv \text{Pr}(N)/\binom{N+M-1}{N}$ and $\text{Pr}(N)$ is the probability of obtaining $N$ photons from $p_U(x)$.

%We consider two different intermediate-size circuits that Ref.~\cite{madsen2022quantum} used to verify their validity again mock-up distributions.
%Note that the largest-scale case, which is claimed to show the quantum advantage, is computationally demanding to compute their output probabilities.
%This is the same reason that Ref.~\cite{madsen2022quantum} used the intermediate-scale circuits.

We emphasize that the XEB we employ is applied for each photon number sector instead of the entire sample set, consistently with the previous methods in experiments \cite{zhong2020quantum, zhong2021phase, madsen2022quantum}.
Therefore, it is not sensitive to the total photon number distribution.
If we conduct the benchmarking for the entire set, a trivial way can spoof it by manipulating the total photon number distribution of a mock-up sampler so that the sampler generates samples from a particular sector having a larger probability of each outcome than other sectors on average.
Meanwhile, we show in Ref.~\cite{supple} that we can adjust our spoofer's total number distribution to be consistent with the ideal case.

Figure~\ref{fig:xanadu} again shows that without post-selection, the XE is much smaller than the experimental samples' XE from Ref.~\cite{madsen2022quantum}.
Meanwhile, for $k=10$, the spoofer's XE is much larger than the experiment for the wide range of photon numbers.
Additionally, as the post-selection rate increases further, the difference becomes more significant.
Based on the trend, we expect that such a gap may not close for the large photon number sectors.
Similar results for different experiments \cite{zhong2021phase} are provided in Ref.~\cite{supple}.

\emph{Prediction for larger systems.---}
Since XEB is computationally inefficient, it is demanding to check if the spoofer works even for larger systems.
%, such as the experiments in Refs.~\cite{zhong2020quantum, zhong2021phase, madsen2022quantum}.
To predict its performance for large-size circuits and how the post-selection overhead scales, we consider a particular type of FS~\footnote{While the Fermionic sampling we consider is easy, there exists different types of Fermionic sampling which is proven to be hard under plausible assumptions~\cite{oszmaniec2022fermion}}, a variant of FBS (See Refs.~\cite{aaronson2013bosonsampling, supple}).
Since the FS is easy to simulate and its probabilities are easy to compute \cite{aaronson2013bosonsampling}, it does not provide quantum advantage.
Nevertheless, its similar structure to BS may enable us to predict a larger BS, since the key idea of our algorithm is to post-select heavy outcomes without computing the probability of the ideal distribution; thus it does not directly rely on the easiness of FS.

\begin{figure}[t]
\includegraphics[width=200px]{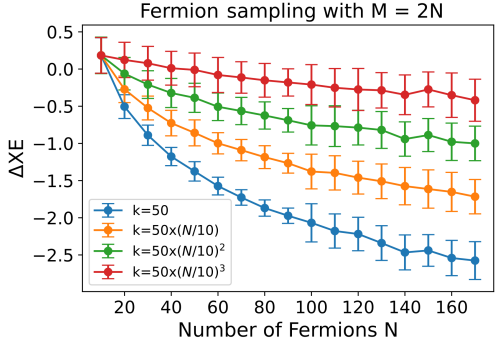}
\caption{XE of FS. $N_s=10^3$ for $N\leq 90$ and $N_s=10^2$ for $N\geq 10^2$. We used $10^3$ different random unitary matrices for circuit configuration. The error bar represents the standard deviation of XE difference.
}
\label{fig:fermionic}
\end{figure}

Figure~\ref{fig:fermionic} shows the result.
We implement the same spoofer for FS with larger system sizes with post-selection rate $k$ differently to analyze the post-selection overhead and investigate the XE difference: $\Delta\text{XE}\equiv \text{XE}_{\text{spoofer}}-\text{XE}_{\text{id}}$.
%First, for constant $k$, as the system size grows, the ideal XE quickly outperforms the spoofer's XE.
%To overcome this, we also increase post-selection rate as the system size.
Although the XE from different choices of the post-selection rate varies, all the different cases' XE difference gradually becomes negative as the system size grows.
Such a trend may be caused because the first-order marginals' correlation to the ideal distribution is not sufficiently large for larger system sizes.
It implies that as the system size increases, the post-selection rate might have to increase, superpolynomially at worst.
Even if this is true, since we are comparing the XE score against the {\it ideal} case and noise significantly decreases the XE of experiments (see below), it would be extremely difficult to surpass the score from our spoofer in the near future.
%Still, analytical prediction of the post-selection overhead is an important open question to determine the efficiency of our algorithm and, consequently hardness of heavy outcome generation of BS.

\emph{Evidence of efficiency of spoofing noisy BS.---}
We now provide evidence that our algorithm can efficiently spoof {\it noisy} BS experiments.
To this end, we study the {\it linear} XE, $\XE\equiv \sum_x q_U(x)p_U(x)$, of ideal and noisy FBS, assuming a collision-free regime, i.e., $M=\omega(N^5)$.
For the ideal case, i.e., ${q_U(x)=p_U(x)=|\per U_x|^2}$ \cite{aaronson2011computational}, the average XE for Haar-random unitary $U$ is given by
\begin{align}
    \mathbb{E}_U[\XE_\text{id}]
    &=\sum_x \mathbb{E}_U[|\per U_x|^4]
    \approx \frac{\sum_x \mathbb{E}_Z[|\per Z|^4]}{M^{2N}}
    \approx\frac{(N+1)!}{M^N}.
\end{align}
Here, we approximated submatrices of large Haar-random unitary matrices by random Gaussian matrices, $U_x\approx Z/\sqrt{M}$ with $Z$ following the complex normal distribution and used $\mathbb{E}_Z[|\per Z|^4]=N!(N+1)!$ \cite{aaronson2011computational} and $\binom{M}{N}\approx M^N/N!$.
Meanwhile, the XE of $q_U$ independent of $p_U$ is given by $\mathbb{E}_U[\XE_\text{idp}]=N!/M^N$ \cite{supple}, where one can clearly see the additional factor $(N+1)$ for the ideal case.

\iffalse
Meanwhile, the XE of $q_U$ independent of $p_U$ is given by \cite{supple}
\begin{align}
    \langle\XE_\text{idp}\rangle_U
    &=\sum_x \langle p_U(x)\rangle_U\langle q_U(x)\rangle_U
    =\frac{N!}{M^N},
\end{align}
where we used $\langle |\per Z|^2\rangle_Z=N!$ \cite{aaronson2011computational}.
One can clearly see the additional factor $(N+1)$ for the ideal case.
\fi

We now analyze the behavior of the XE under partial distinguishability \cite{tichy2015sampling, renema2018efficient, supple}, one of the most important noise models in practice, describing partial overlaps of wave functions of different photons (See Refs.~\cite{tichy2015sampling, renema2018efficient, supple} for more details of the model.)
The XE of BS under partial distinguishability $0\leq \rho <1$ with respect to the ideal distribution, normalized by the XE of an independent distribution, is bounded by \cite{supple}
\begin{align}\label{eq:XE_pd}
    \frac{\mathbb{E}_U[\text{XE}_\text{pd}]}{\mathbb{E}_U [\text{XE}_\text{idp}]}\leq \frac{e^2(1-\rho^{N+1})}{1-\rho}.
\end{align}
Observe that even for constant $\rho$, the normalized XE converges to ${e^2/(1-\rho)}$, which only provides an additional constant factor $e^2/(1-\rho)$, while the ideal score's factor increases as $(N+1)$.~Clearly, the XE significantly decreases under experimental noise, suggesting that the experimental XE will be much smaller than the ideal case.

\iffalse
We now analyze the behavior of the XE under loss with a transmission rate $\eta$.
Then the probability of $N$-photon outcome $x$ with loss becomes $q_U(x)=\eta^N p_U(x)$.
Thus, the linear XE between them is written as
\begin{align}
    \langle \XE_\text{loss}\rangle_U
    =\eta^N \langle\XE_\text{id}\rangle_U
    \approx \frac{\eta^N(N+1)!}{M^N},
\end{align}
which shows that the XE decreases exponentially in $\eta$.

Even for constant $\eta$, the XE becomes exponentially small, although this regime is already hard to implement in practice since we need to maintain a constant transmission rate with scaling up.
Here, the XE under loss can be smaller than the independent case because the probability of obtaining $N$-photon outcomes decreases.
\fi

%We now provide analytical evidence supporting our numerical results for FBS case with a function $h_U(x)$ correlated with the ideal distribution $p_U(x)$.
We now provide analytical evidence that the noisy XE might be attainable using our method.
Since analyzing our algorithm's output probability distribution is difficult due to post-selection, we consider a different probability distribution $\propto [h_U(x)]^s$, which contains our algorithm's core idea as it tends to generate heavy outcomes from $p_U(x)$.
Effectively, a similar effect to post-selection is expected for $s>1$, because the resulting probability favors large-probability outcomes from $h_U(x)$ (we do not expect that we can sample from the distribution.).
Thus, the power $s$ is associated with post-selection overhead $k$.
We show for linear XE that the distribution with a multinomial distribution $h_U(x)=N^{-N}\prod_{i=1}^N\sum_{j=1}^N |U_{j,x_i}|^2$,
provides \cite{supple}
\begin{align}
    \frac{\mathbb{E}_U[\text{XE}(p_U,h^s_U)]}{\mathbb{E}_U[\text{XE}_\text{idp}]}
    &\equiv \frac{1}{\mathbb{E}_U[\text{XE}_\text{idp}]}\mathbb{E}_U\left[\frac{\sum_{x} p_U(x)h^s_U(x)}{\sum_{x} h^s_U(x)}\right]
    \approx e^s,
\end{align}
where the approximation holds for small $s$ and large $N$.
Since it suffices to achieve a constant factor due to noise from Eq.~\eqref{eq:XE_pd} and there are other types of noise such as loss \cite{supple}, we expect that choosing a constant power $s$, which may be interpreted as constant post-selection overhead $k$, might be sufficient to spoof noisy BS unless the noise can be highly suppressed.

\iffalse
We consider Fock-state boson sampling with $N$ photons and a linear-optical circuit $U$, whose probability distribution is written as
\begin{align}
    p_U(\bm{x})=\frac{|\per U_{N,\bm{x}}|^2}{\bm{x}!},
\end{align}
and mock-up distribution $h_U(x)$ as a 
\begin{align}
    h_U(\bm{x})=\frac{1}{N^N}\prod_{i=1}^N\sum_{j=1}^N |U_{j,x_i}|^2.
\end{align}
\fi

\emph{Comparison to existing spoofing methods.---}\label{sec:diff}
Most existing methods for spoofing GBS work by attempting to model noisy experiments. 
It is often the case that noisy experiments converge to classically easy distribution \cite{kalai2014gaussian, renema2018classical, renema2018efficient, oh2021classical, villalonga2021efficient, qi2020regimes, martnez2022classical}.
These algorithms exploit this by sampling from this easy distribution. 
For example, G.~Kalai claimed that noisy experimental boson sampler’s probabilities are approximable by low-degree polynomials \cite{kalai2014gaussian}.
Since then, there have been many subsequent proposals to take advantage of noise \cite{renema2018efficient, renema2018classical}.
Similarly, the algorithm in Ref.~\cite{villalonga2021efficient} aims to reproduce low-order marginals without recovering high-order marginals. 
Another method is to use the classical state because noise often transforms the output state to be close to classical states~\cite{oszmaniec2018classical, garcia2019simulating, qi2020regimes, martnez2022classical}.

Unlike the existing methods, our method's goal is spoofing XEB instead of approximate simulation.
Since XE increases by generating heavy outcomes, a large XE can be achieved without simulation.
Thus, our approach does not necessarily work for other benchmarking that does not rely on heavy outcome generation, such as Bayesian test and correlation functions \cite{phillips2019benchmarking}.
The Bayesian test employed in current GBS experiments as another evidence of quantum advantage \cite{zhong2020quantum, zhong2021phase, madsen2022quantum} defines the score as
\begin{align}
    \text{score}=\frac{1}{N_s}\sum_{i=1}^{N_s}\log\frac{p_U(x^{(i)})}{q_U(x^{(i)})},
\end{align}
where $\{x^{(i)}\}_{i=1}^{N_s}$ is a sample set from {\it experiment}.
A positive score implies that the experimental samples are more likely to be sampled from the ideal distribution $p_U(x)$ than the mock-up distribution $q_U(x)$.
Because the test requires the profile of the mock-up distribution and computing the post-selected distribution's probability is difficult, it is not easy to perform Bayesian test against our method.
Nevertheless, because our method favorably generates heavy outcomes, the output distribution may be highly concentrated.
Thus, even if we can compute the probability distribution, our method is unlikely to pass the Bayesian test.

\begin{figure}[t]
\includegraphics[width=180px]{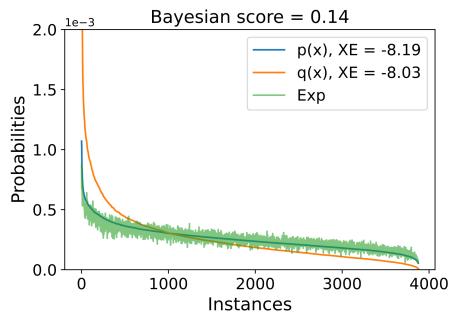}
\caption{Bayesian test using $M=16$ and $N=4$ case in Ref.~\cite{madsen2022quantum}. 
The green curve is the experimental data from Ref.~\cite{madsen2022quantum}. 
The XE score is different from Fig.~\ref{fig:xanadu3} because the probability is normalized over the sector for this figure.}
\label{fig:bayesian}
\end{figure}

As an example that generates heavy outcomes but fails to pass Bayesian test, consider a mock-up distribution $q_U(x)\propto p_U(x)^2$ (without post-selection).
Thus, $q_U(x)$ generates heavier outcomes than $p_U(x)$ with respect to $p_U(x)$, which guarantees to spoof the XE test, as shown in Fig.~\ref{fig:bayesian}.
However, since the mock-up distribution is farther than the ideal distribution to empirical experimental distribution, the Bayesian score becomes positive, implying that $q_U(x)$ fails to pass the Bayesian test although it generates heavy outcomes.
%Here we have used the experimental samples from Ref.~\cite{madsen2022quantum} for Bayesian test.
%Hence, it would be another interesting open question to find a method to spoof the Bayesian test and even both tests at the same time or to show that spoofing Bayesian test is hard.

%\emph{Conclusion.---}\label{sec:conclusion}
%We have studied XE score in FBS and proposed a classical algorithm that can generate heavy outcomes of BS and obtain a large XE score without simulating the ideal sampler.
%We have numerically demonstrated that our spoofer can attain a larger cross entropy score than even the ideal probability distribution for a small system size and that its cross entropy score for intermediate-scale boson sampler is larger than the experiments.
Finally, our results spoofing the XE test do not imply that the experiments in Refs.~\cite{zhong2020quantum, zhong2021phase, madsen2022quantum} are easy to simulate because our algorithm's goal is to spoof the test without simulation. Instead, the implication is that the XE scores may not be a proper measure as evidence of quantum advantage of BS.
Our results open many questions about the verification of sampling tasks.
First, making our method analytical is crucial to predict the asymptotic behavior of the method precisely.
Second, it would also be interesting to apply the same method to other sampling tasks.
Especially for RCS, finding a distribution that is correlated with the ideal distribution and easy to compute would be an important first step to applying the presented method.
We emphasize again that although we chose $h_U$ relying on marginals, other various possible quantities may not necessarily rely on marginals.
Also, since our algorithm is specialized to spoof the XE test, we do not expect it to pass other tests such as the Bayesian test.
Hence, it would also be interesting to analyze the Bayesian test as well to see if it can be spoofed.

\begin{acknowledgements}
We thank Benjamin Villalonga and Sergio Boixo for interesting and fruitful discussions.
We acknowledge support from the ARO MURI (W911NF-16-1-0349, W911NF-21-1-0325), AFOSR MURI (FA9550-19-1-0399, FA9550-21-1-0209), DoE Q-NEXT, NSF (OMA-1936118, ERC-1941583, OMA-2137642), NTT Research, and the Packard Foundation (2020-71479).
B.F. acknowledges support from AFOSR (FA9550-21-1-0008). This
material is based upon work partially
supported by the National Science Foundation under Grant CCF-2044923
(CAREER) and by the U.S. Department of Energy, Office of Science,
National Quantum Information Science Research Centers (Q-NEXT).
This material is also supported in part by the DOE QuantISED grant DE-SC0020360.
The authors are also grateful for the support of the University of Chicago Research Computing Center for assistance with the numerical simulations carried out in this paper.
\end{acknowledgements}

\bibliography{reference}

\end{document}

\definecolor{navy}{RGB}{46,72,102}
\definecolor{pink}{RGB}{219,48,122}
\definecolor{grey}{RGB}{184,184,184}
\definecolor{yellow}{RGB}{255,192,0}
\definecolor{grey1}{RGB}{217,217,217}
\definecolor{grey2}{RGB}{166,166,166}
\definecolor{grey3}{RGB}{89,89,89}
\definecolor{red}{RGB}{255,0,0}

\preprint{APS/123-QED}

\title{Supplemental Material: Spoofing cross entropy measure in boson sampling}
\author{Changhun Oh}
%\email{changhun@uchicago.edu}
\affiliation{Pritzker School of Molecular Engineering, University of Chicago, Chicago, Illinois 60637, USA}
\author{Liang Jiang}
\affiliation{Pritzker School of Molecular Engineering, University of Chicago, Chicago, Illinois 60637, USA}
\author{Bill Fefferman}
\affiliation{Department of Computer Science, University of Chicago, Chicago, Illinois 60637, USA}

\maketitle

\renewcommand{\thesection}{S\arabic{section}}
\renewcommand{\theequation}{S\arabic{equation}}
\renewcommand{\thefigure}{S\arabic{figure}}

\tableofcontents

%\newpage

\section{Spoofing USTC experiment}
In this section, we apply our spoofing algorithm to another experiment in Ref.~\cite{zhong2021phase}.
Note that one difference of the experiment in Ref.~\cite{zhong2021phase} from the one in Ref.~\cite{madsen2022quantum} is that the former used a threshold detector instead of a photon number resolving detector.
While there were many attempts to spoof the experiment \cite{kalai2014gaussian, renema2018classical, renema2018efficient, villalonga2021efficient, oh2022classical, popova2022cracking, martnez2022classical}, the most related strategy is to use a probability distribution which has the same low-order marginals by greedy algorithm \cite{villalonga2021efficient}.
For different measures such as total variation distance or Kullback-Leibler divergence, the low-order marginal approximation outperformed for wide range of region, while it was inconclusive for the cross-entropy (XE) measure since each score is within the error bar.
Also, since the total variation distance and Kullback-Leibler divergence were used only for marginals due to the computational cost, and as pointed out in Ref.~\cite{madsen2022quantum}, the behavior of the score in marginal modes might not be consistent with the full distribution's one.
On the other hand, for the XE measure, Ref.~\cite{villalonga2021efficient} indeed used the full distribution with fixed photon click outcomes.

We now compare the greedy algorithm from Ref.~\cite{villalonga2021efficient} with the proposed spoofer in the present work.
Notice that while Ref.~\cite{villalonga2021efficient} uses up to third-order marginals, we only use up to second-order marginals.
Nevertheless, the difference between the scores from the two distributions was not considerable \cite{villalonga2021efficient}.
Also, we note that we normalized the XE measure by the probability of a fixed sector of photon clicks.
We selected three intermediate-scale Gaussian boson sampling (GBS) experiments in Ref.~\cite{zhong2021phase}.
The system parameters of each example are 
(a) power $P=0.5$, beam waist $125\mu\text{m}$, squeezing parameter around $r\approx 0.515$, and the ideal mean click number $7.27$, 
(b) power $P=0.15$, beam waist $65\mu\text{m}$, squeezing parameter around $r\approx 0.495$, and the ideal mean click number $5.98$, and 
(c) power $P=0.3$, beam waist $65\mu\text{m}$, squeezing parameter around $r\approx 0.692$, and the ideal mean click number $11.94$.
Note that a similar approach did not succeed to attain a higher XE than that of the experimental samples in Ref.~\cite{madsen2022quantum}.

\begin{figure*}[t]
\includegraphics[width=500px]{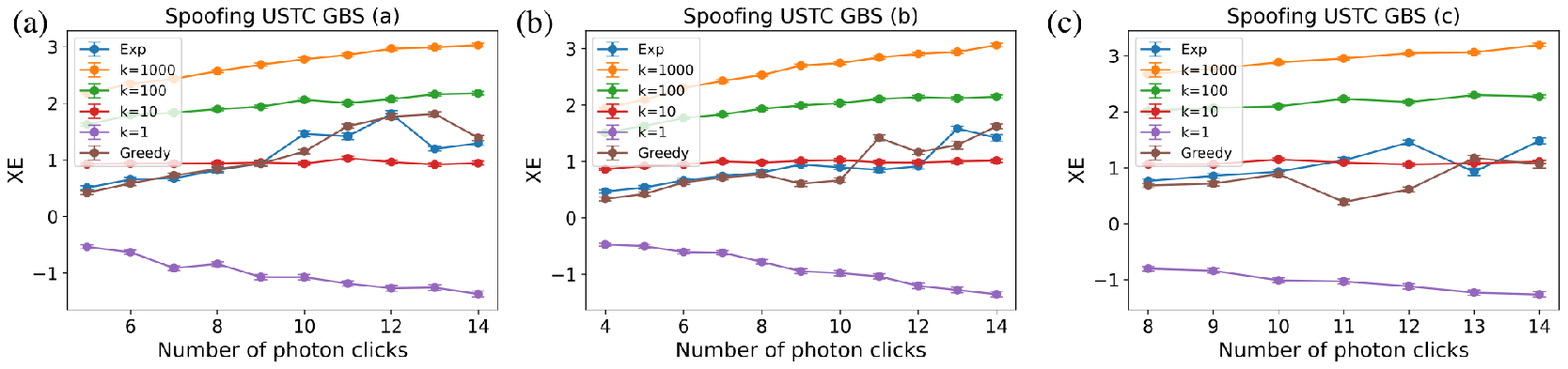}
\caption{Spoofing the small-size Gaussian boson sampling experiments in Ref.~\cite{zhong2021phase}. Note that the ideal mean click number is $\bar{N}\approx 7.27, 5.98, 11.94$, respectively.}
\label{fig:USTC}
\end{figure*}

Our results are shown in Fig.~\ref{fig:USTC}.
First of all, as we observed from the previous examples, the first-order-marginal-based distribution $(k=1)$ does not provide a large XE.
Again, as we post-select with post-selection rates such as $k=100,1000$, the score is much larger than both experimental samples and the samples from the greedy algorithm.
It is worth emphasizing that for experiments (a) and (b), the number of photon clicks we focus on ranges the dominant photon number outcomes and that we expect that such intermediate-scale experiments are less affected by experimental noises.
In addition, as the number of photon clicks grows, the increment does not have a significant difference between the experimental XE or the existing spoofer and our spoofer.
Therefore, such results show strong evidence of spoofing with respect to XE measure and also much stronger evidence of spoofing than Ref.~\cite{villalonga2021efficient}.

\section{Discussion about choice of $q(x)$ and $h(x)$}
\subsection{Choice of sampler $q(x)$}\label{sec:q}
In general, $q(x)$ in our spoofing algorithm does not necessarily correlate with the ideal probability $p$.
Thus, we can simply choose $q(x)$ to be uniform distribution although it is obviously advantageous to choose $q(x)$ to be correlated with the ideal distribution to boost the post-selection procedure.
Thus, to fully exploit this fact, one may choose $q(x)$ to be a probability distribution that is more correlated to $p(x)$ even if its probability cannot be efficiently computed, for example thermal state approximation \cite{garcia2019simulating}, closest classical state \cite{qi2020regimes, martnez2022classical}, and second-order-marginal-based distribution \cite{villalonga2021efficient}.

Meanwhile, we emphasize that depending on the choice of the sampler $q(x)$, the total photon number distribution may vary and deviate from the ideal distribution in general.
For example, for certain samplers $q(x)$, the probability of obtaining samples of a high total photon number can be significantly smaller than the ideal one, so that there might be a large overhead to generate samples of large total photon numbers.
A possible remedy is to introduce rejection sampling with respect to the ideal total photon number distribution while it is still possible that the rejection sampling is highly inefficient when the total photon number distributions are very different.
The simplest way to address the issue is to choose $q(x)$ to be the uniform distribution for each photon number sector and then output samples with matching the total photon number distribution.
For this reason and simplicity, we choose uniform distribution as a sampler $q(x)$ throughout the present work.
Thus, in principle, we can generate the correct total photon number distribution by simply adjusting the number of samples for each photon number sector.
Note that the total photon number distribution can be efficiently computed using convolution \cite{hamilton2017gaussian, deshpande2021quantum} and that the total photon number click distribution, corresponding to threshold detectors, can be efficiently approximated using the method in Ref.~\cite{drummond2022simulating}.
Another possible choice is the distinguishable boson sampler, which provides the same total photon number distribution as the ideal one (but not necessarily the same total photon click distribution).
%In the next section, we numerically demonstrate that our algorithm can generate heavy outcomes and so obtain a large cross entropy score and spoof the most recent Gaussian boson sampling experiments for cross entropy benchmarking.

\subsection{Choice of $h(x)$}
Although there can be various candidates of the correlated distribution $h(x)$, in the main text, we focused on a probability distribution that only relies on the first-order marginals of the ideal distribution, which can be used for both Fock-state BS (FBS) and GBS.
We again emphasize that $h(x)$ does not need to be a probability distribution because its important property is the correlation with the ideal probability distribution $p(x)$.
Specifically, to construct $h(x)$, we first compute the first-order marginals of all $M$ modes, i.e., $p_1(x_1),\dots,p_M(x_M)$, where $p_i(x_i)$ represents the $i$th mode's marginal probability of outcome $x_i$.
Then, the first-order-marginal-based distribution is obtained simply by the product of the marginals, i.e.,
\begin{align}
    h(x)\equiv \prod_{i=1}^M p_i(x_i).
\end{align}
Since it correctly reproduces the first-order marginals by definition, it contains a correlation with the ideal distribution $p(x)$ \cite{villalonga2021efficient}.
In addition, the probability of each outcome can be efficiently computed.

For FBS, we provide two additional examples of such correlated distributions which can be efficiently computed in Sec.~\ref{app:mock-up}.
An important implication of these examples is that our scheme does not require a correlation specifically from marginals and that there can be other quantities that do not directly rely on the marginals.
This is crucial to extend our method to RCS where it is known that the reduced density matrix of small number of qubits out of large number of qubits is exponentially close to the maximally mixed state, which indicates that marginal-based mock-up distribution is not a good candidate of the correlated distribution to implement our method for random circuit sampling.

\section{Mock-up distributions for Fock-state boson sampling}\label{app:mock-up}
In this section, we provide two additional correlated distributions that can be used to generate heavy outcomes as well as the first-order-marginal-based algorithm for FBS.
These additional distributions highlight that they do not necessarily rely on marginals to generate heavy outcomes.
%Note that these distributions can be used as a mock-up sampler $q(x)$ for the first algorithm or a correlation indicator $h(x)$ for the second algorithm.
For simplicity, we will call these correlated distributions.

Before introducing those two distributions, we briefly note the implementation of the first-order-marginal-based algorithm for FBS.
In contrast to GBS, where the reduced density matrix for each mode is again a Gaussian state so the marginal probability is straightforward to compute, for FBS, it is not very straightforward.
Nevertheless, Refs.~\cite{aaronson2011computational, ivanov2020complexity} provide the method of computing marginal probabilities of a small number of modes efficiently.
Note that a similar type of samplers have been studied in different contexts~\cite{tichy2014stringent, oszmaniec2018classical}.

The first multinomial algorithm, which we call the multinomial algorithm with uniform mixed input (the meaning will be clear from below), is motivated in Ref.~\cite{aaronson2013bosonsampling}, where an efficient measure of discriminating a boson sampler from a uniform sample is provided and a spoofer of that particular measure is introduced.
The correlated distribution is obtained by multinomial distribution of probabilities
\begin{align}
    h_i=\frac{1}{N}\sum_{j=1}^N |U_{ji}|^2~~~\text{for}~~~ i \in [M].
\end{align}
Intuitively, this probability can be thought of as an output probability after injecting a single photon randomly on one of the first $N$ modes for $N$ times, i.e.,
\begin{align}
    \hat{\rho}=\frac{1}{N}\sum_{i=1}^N |\psi_i\rangle\langle \psi_i|,
\end{align}
where $|\psi_i\rangle$ is a single-photon state on $i$th mode.

The second multinomial algorithm, which we call the multinomial algorithm with equal superposition input, is to use an input state that is an equal superposition of a single-photon in $N$ input modes, i.e., 
\begin{align}
    \frac{1}{\sqrt{N}}(|10\dots0\rangle+|010\dots 0\rangle+\dots+ |0\dots 01\rangle)\otimes |0\dots0\rangle.
\end{align}
The difference from the previous method is that now the single photon has a superposition over the $N$ modes instead of being a statistical mixture.
Since each trial we only get a single-photon click, we repeat this for $N$ times to obtain $N$ photon outcomes.
Similarly to the previous algorithm, it can be described by a multinomial distribution with
\begin{align}
    h_i=\frac{1}{N}\sum_{j,k=1}^N U_{ki}U_{ji}^*~~~\text{for}~~~ i \in [M].
\end{align}

\begin{figure*}[t]
\includegraphics[width=500px]{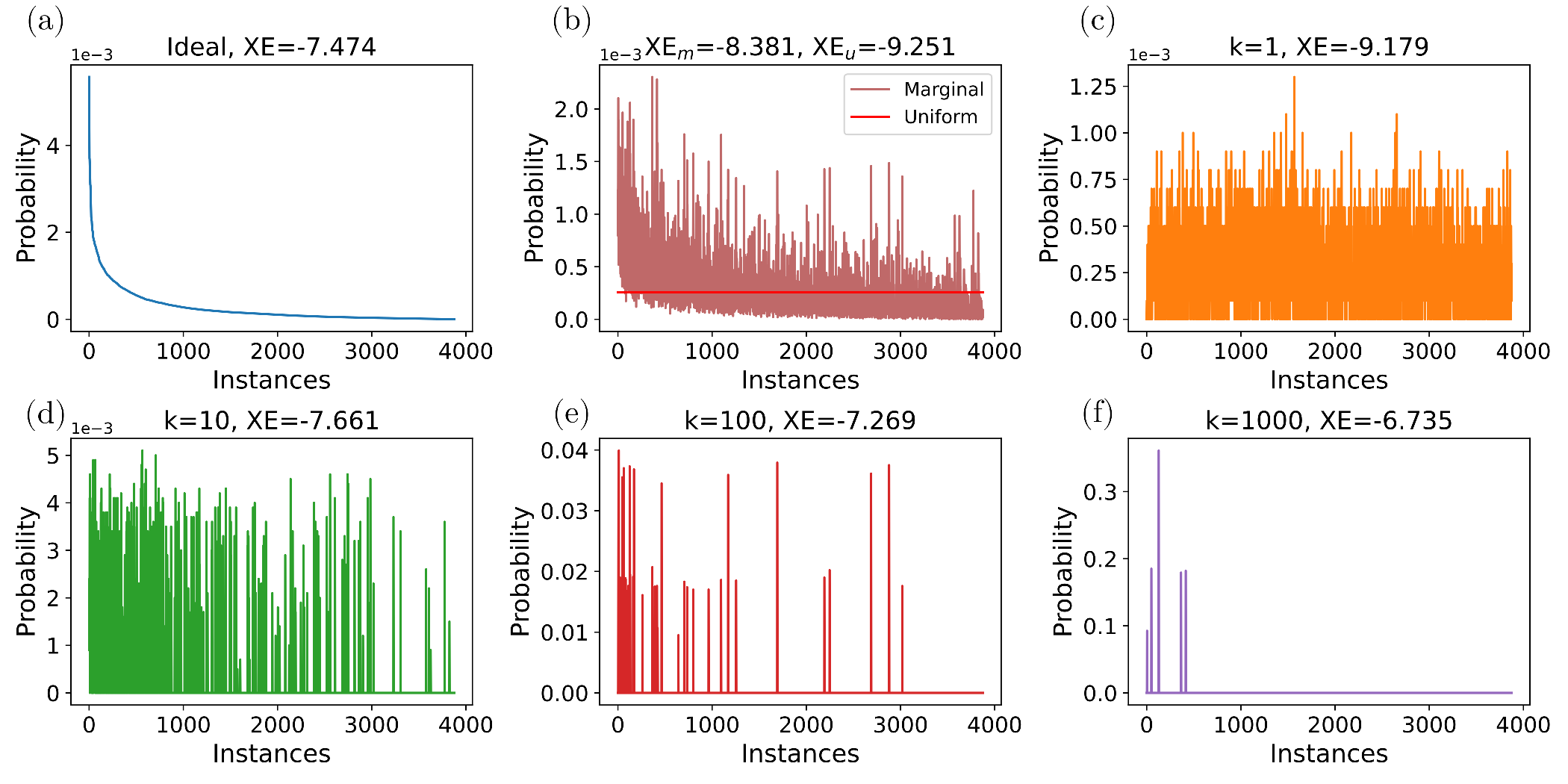}
\caption{16 modes FBS case with 4 particles using the first-order-marginal-based method. 
(a) Ideal distribution is obtained by directly computing individual probabilities. 
(b) The correlated distribution $h(x)$ based on the first-order marginal is also obtained by directly computing all probabilities.
(c)-(f) The spoofing method with different $K=1,10,100,1000$.
For $k=1$, we do not post-select, so it is an empirical distribution of the uniform distribution.
The number of samples for (c)-(f) is $N_s=10^4$.
}
\label{fig:bosonic3_marginal}
\end{figure*}

\begin{figure*}[t]
\includegraphics[width=500px]{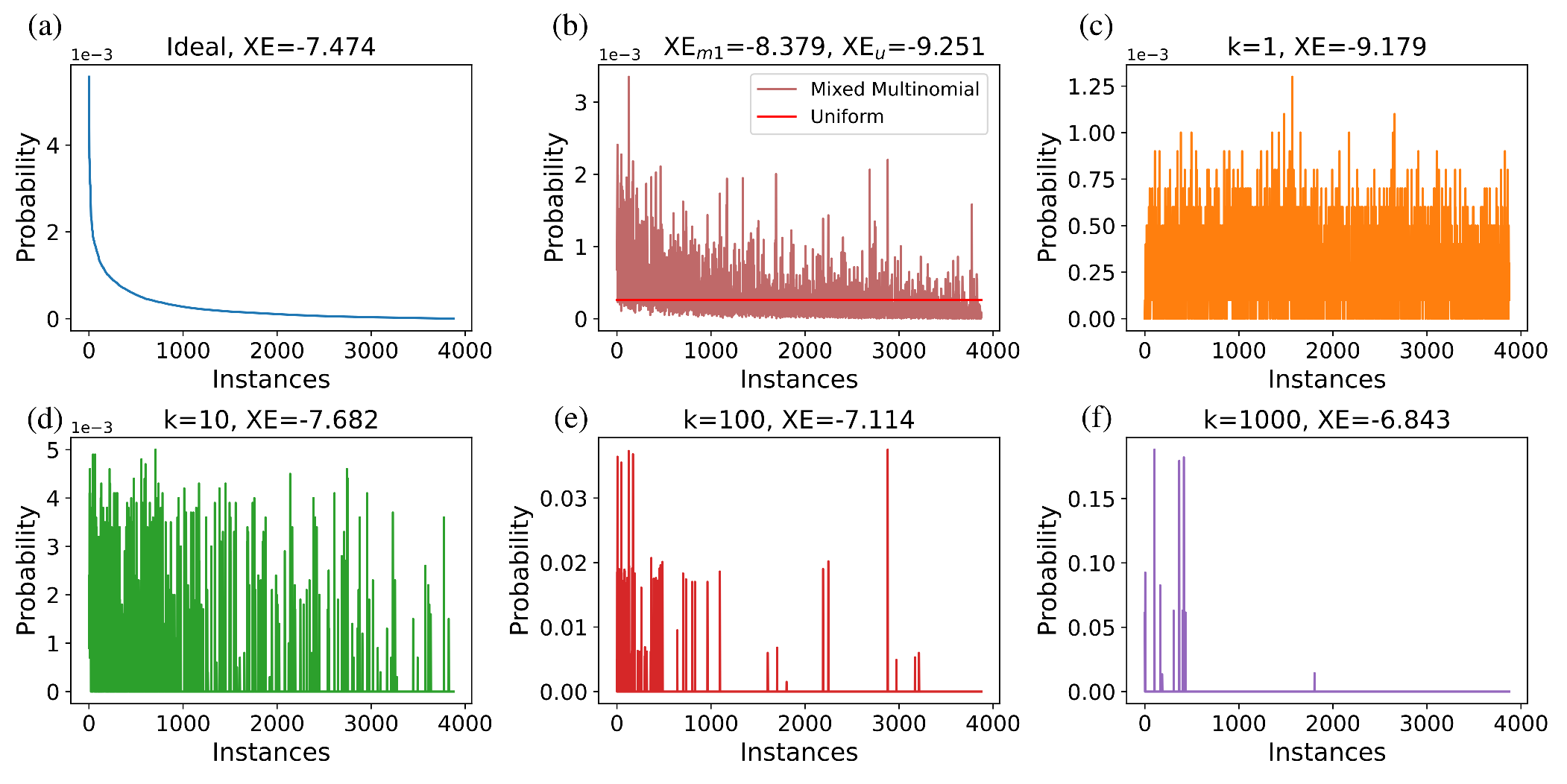}
\caption{16 modes FBS case with 4 particles using the multinomial algorithm with uniform mixed input. 
(a) Ideal distribution is obtained by directly computing individual probabilities. 
(b) The correlated distribution $h(x)$ based on the multinomial is also obtained by directly computing all probabilities.
(c)-(f) The spoofing method with different $K=1,10,100,1000$.
For $k=1$, we do not post-select, so it is an empirical distribution of the uniform distribution.
The number of samples for (c)-(f) is $N_s=10^4$.
}
\label{fig:bosonic3}
\end{figure*}

\begin{figure*}[t]
\includegraphics[width=500px]{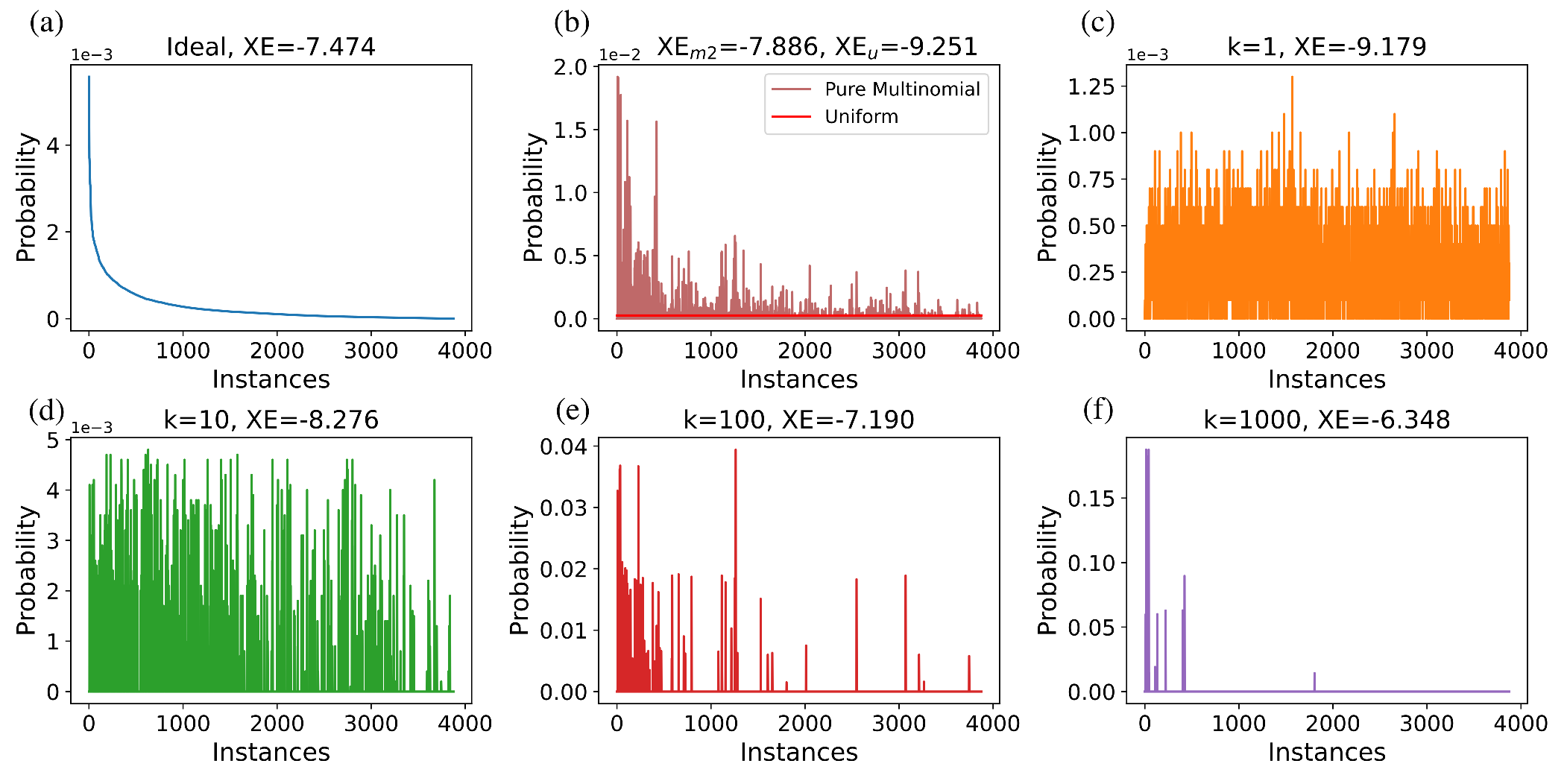}
\caption{16 modes FBS case with 4 particles using the multinomial algorithm with equal superposition input.
(a) Ideal distribution is obtained by directly computing individual probabilities. 
(b) The correlated distribution $h(x)$ based on the multinomial is also obtained by directly computing all probabilities.
(c)-(f) The spoofing method with different $K=1,10,100,1000$.
For $k=1$, we do not post-select, so it is an empirical distribution of the uniform distribution.
The number of samples for (c)-(f) is $N_s=10^4$.
}
\label{fig:bosonic3-2}
\end{figure*}

Using the same post-selection procedure with the uniform sampler and those three different correlated distributions, we analyze the case with $M=16$ and $N=4$ in Figs.~\ref{fig:bosonic3_marginal},\ref{fig:bosonic3} and \ref{fig:bosonic3-2}.
One can easily see that all the three schemes can be used to post-select heavy outcomes.
Especially, the multinomial algorithm with equal superposition input has a stronger correlation with the ideal probability distribution than other algorithms including the marginal-based distribution, which suggests that there can be other mock-up distributions that have a better correlation than the marginal-based algorithm.

\section{Fock-state boson sampling and fermion sampling}\label{app:fermionic}
In this section, we briefly recall the basics of FBS \cite{aaronson2011computational} and FS \cite{aaronson2013bosonsampling} for completeness.
First, for FBS we first prepare $N$ single-photon input state, $|1\dots,1,0,\dots,0\rangle$ and let the input state go through an $M$-mode linear-optical circuit, and measure the output photon configuration.
We can then show that the output probability is written as
\begin{align}
    p(x)=\frac{|\per(U_{x})|^2}{x!},
\end{align}
where $U_{x}$ is the submatrix of unitary $U$ with selecting the first $N$ rows and $i$th columns repeating $x_i$ times.
By using the (conjectured) hardness of approximating the permanent of Gaussian random matrices, approximate boson sampling can be proven to be hard under some plausible assumptions (see Ref.~\cite{aaronson2011computational} for a more rigorous argument).

On the other hand, for FS, instead of using photons, which follow bosonic statistics, we use $N$ fermions.
The most important difference between two types of particles is that a single mode cannot be occupied by more than a single particle for fermions, whereas it can be for bosons.
Such a difference changes the probability distribution as
\begin{align}
    p(x)=|\text{Det}(U_{x})|^2,
\end{align}
where $x$ is now a bit string because a single mode can only be in a vacuum or a single-fermion state.
The consequence of the different statistics also changes the probability from permanent, which is hard to compute, to determinant, which is easy to compute.

\begin{figure*}[t]
\includegraphics[width=500px]{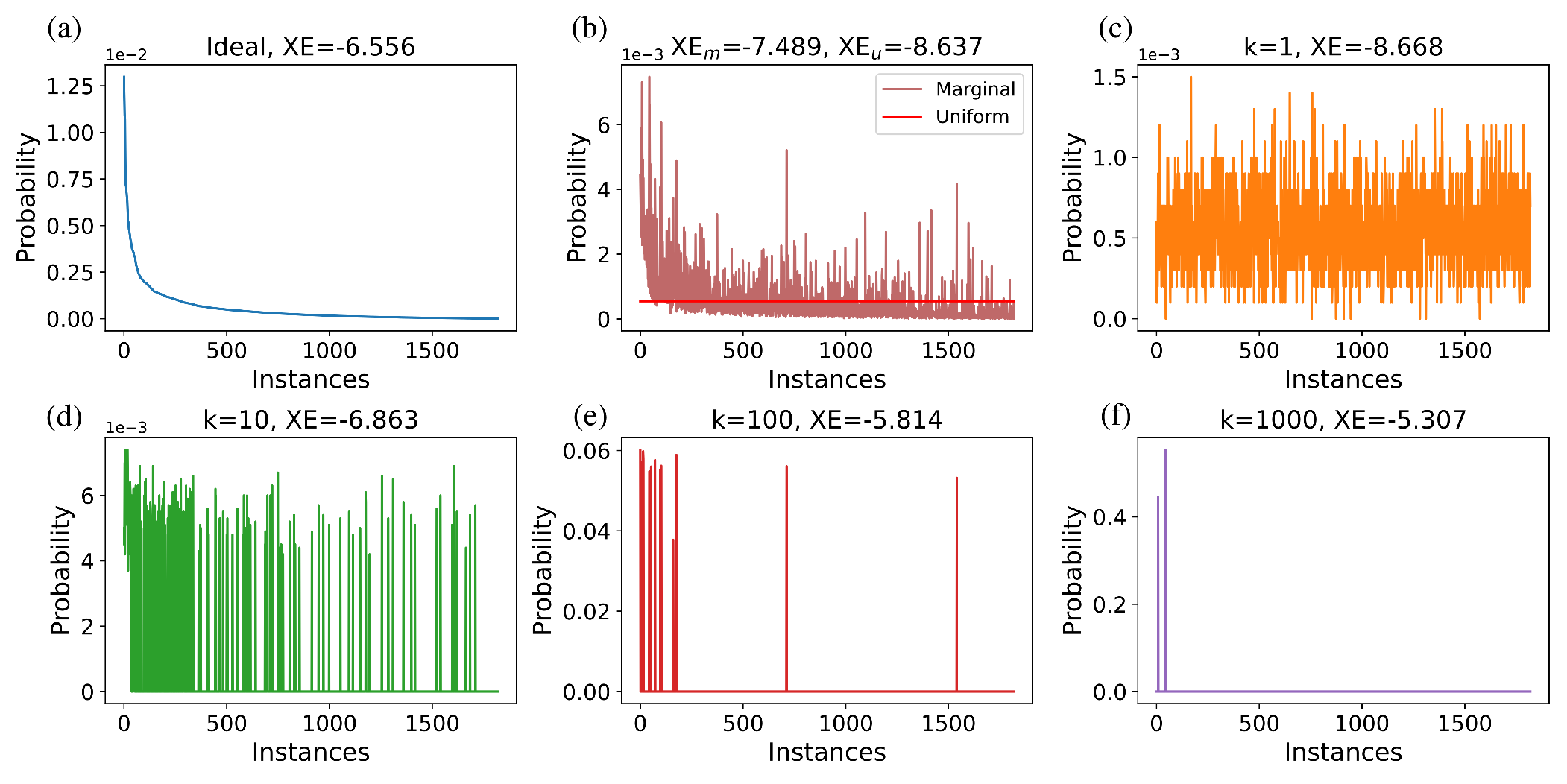}
\caption{16 modes fermion sampling case with 4 particles. 
(a) Ideal distribution is obtained by directly computing individual probabilities. 
(b) The probability distribution based on the first-order marginal is also obtained by directly computing all probabilities.
(c)-(f) The spoofing method with different $K=1,10,100,1000$.
For $k=1$, we do not post-select, so that it is an empirical distribution from (b).
The number of samples for (c)-(f) is $N_s=10^4$.
}
\label{fig:fermionic3}
\end{figure*}

For comparison, we implement our algorithm for $M=16$ modes and $N=4$ particle case for FS.
The result is shown in Fig.~\ref{fig:fermionic3}.

\section{XE score for ideal and noisy FBS}
\subsection{Ideal score}
We study the XE measure in FBS \cite{aaronson2011computational}.
Let us first consider collision-free FBS, i.e., $M=\omega(N^5)$, which is the standard case of BS.
In this case, we can assume that the relevant matrix is described by a random Gaussian matrix.
Let $p_U(x)$ be the ideal FBS probability distribution with $N$ single photons, which is written as
\begin{align}
    p_U(x)=|\per U_x|^2.
\end{align}
The linear XE of a probability distribution $q_U(x)$ with respect to $p_U(x)$ is given by
\begin{align}
    \text{XE}
    =\sum_x p(x)q(x).
\end{align}
When $q(x)=p(x)$, i.e., the ideal case,
\begin{align}
    \XE_\text{id}=\sum_x p(x)^2.
\end{align}
For Haar-random unitary $U$, or more precisely random Gaussian matrix $U_x\approx Z/\sqrt{M}$ with $Z$ following the complex normal distribution,
\begin{align}
    \mathbb{E}_U[\XE_\text{id}]
    &=\sum_x \mathbb{E}_U[p_U(x)^2]
    =\binom{M}{N}\frac{N!(N+1)!}{M^{2N}} 
    \approx \left(\frac{Me}{N}\right)^N\frac{(N+1)!}{M^{2N}}\left(\frac{N}{e}\right)^{N}
    =\frac{(N+1)!}{M^N},
\end{align}
where the sum is over collision-free outcomes $x$.
%Although we have averaged over different unitaries $U$, we expect that the same relation holds for generic Haar-random unitary $U$ because of the random Gaussian matrix nature.
On the other hand, for two independent distributions $p_U$ and $q_U$, we have
\begin{align}
    \mathbb{E}_U[\XE_\text{idp}]
    &=\sum_x \mathbb{E}_U[p_U(x)]\mathbb{E}_U[q_U(x)]
    =\sum_x \frac{N!}{M^N}\mathbb{E}_U[q_U(x)] \nonumber =\frac{N!}{M^N}\mathbb{E}_U\left[\sum_x q_U(x)\right]
    =\frac{N!}{M^N}.
\end{align}

\subsection{Lossy FBS}
We now analyze the behavior of the linear XE for several different noise models such as photon loss and partial distinguishability.
Consider photon-loss with a  {\it transmission rate} $\eta$.
Then the probability of obtaining $N$-photon outcome $x$ without and with loss is given by
\begin{align}
    p(x)=|\per U_x|^2,~~~ q(x)=\eta^N |\per U_x|^2.
\end{align}
Thus, the linear XE between them is written as
\begin{align}
    \mathbb{E}_U[\XE_\text{loss}]=\sum_x \mathbb{E}_U [p(x)q(x)]
    =\eta^N \mathbb{E}_U[\XE_\text{id}]
    =\eta^N\frac{(N+1)!}{M^N}.
\end{align}

\iffalse
Especially when $\eta=1-k/N$ and if we take a limit of $N\to \infty$ to analyze an asymptotic regime,
\begin{align}
    \lim_{N\to\infty}\frac{1}{N}\left[\left(1-\frac{k}{N}\right)^N(N+1)-1\right]
    =\frac{1}{e^k}.
\end{align}
We note that this case is still hard to classically simulate.
This is because boson sampling with a photon loss and boson sampling with a photon loss rate are polynomially equivalent \cite{oszmaniec2018classical}.
Also, $\eta=1-k/N$ case would maintain $N$ photon cases dominantly.
More precisely, the probability that no photon is lost is
\begin{align}
    \eta^N, ~~~ \lim_{N\to \infty}\eta^N=\frac{1}{e^k}.
\end{align}
It means that by using such a lossy boson sampler, we can simulate a lossless boson sampling because the probabilities on the support of $N$-photon space are proportional to each other.
If we choose $\eta$ to be a constant, the normalized XE becomes exponentially small, although this regime is already hard to implement.
\fi

%We can also derive the XE under partial distinguishability \cite{tichy2015sampling, renema2018efficient}, which is provided in the following section.

\subsection{FBS under partial distinguishability}
The output probability after the partial distinguishability noise is written as \cite{tichy2015sampling, renema2018efficient}
\begin{align}
    q_U(x)=\sum_{\sigma\in \mathcal{S}_N}\left(\prod_{j=1}^N S_{j,\sigma(j)}\right)\text{Per}(M*M_{1,\sigma}^*),
\end{align}
where $\mathcal{S}_N$ is the permutation group for $N$ elements, $*$ is the elementwise product and $M_{1,\sigma}$ means that the rows are not permuted and the columns are permuted following $\sigma$.
We assume for simplicity that $S_{ij}(x)=x+(1-x)\delta_{ij}$, which means that any pairs of photons have the same partial distinguishability.
Now, let us rewrite the probability using an orthogonal basis.
Note that $(M*M_{1,\sigma}^*)_{ij}=M_{ij}M^*_{i\sigma(j)}$.
First consider the ideal output probability:
\begin{align}
    p_U(x)=|\text{Per}(Z)|^2=\sum_{\sigma,\rho}\prod_{i=1}^N z_{i,\sigma(i)}z^*_{i,\rho(i)}.
\end{align}
We introduce a basis of polynomials and rewrite the probability:
\begin{align}
    1,h_1(z)\equiv zz^*,h_{2}(z_i,z_j)\equiv z_iz_j^*,~~~z_i\neq z_j^*.
\end{align}
Here, we note that $zz^*$ and $zz^*-1$ do not have any difference for our purposes because $zz^*$ and $1$ have the same eigenvalues.
The effect of partial distinguishability is to add a factor $1, 1, x$ when each polynomial appears in the expression, respectively.
Then, the ideal probability changes under partial distinguishability as
\begin{align}
    p_U(x)
    =|\text{Per}(Z)|^2=\sum_{\sigma,\rho}\prod_{i=1}^N z_{i,\sigma(i)}z^*_{i,\rho(i)}
    \to q_U(x)=\sum_{\sigma,\rho}x^{N-k}\prod_{i=1}^N z_{i,\sigma(i)}z^*_{i,\rho(i)},
\end{align}
where $k$ is the number of $i$'s such that $\sigma(i)=\rho(i)$.
Then,
\begin{align}
    \prod_{i=1}^N z_{i,\sigma(i)}z^*_{i,\rho(i)}
    &=\prod_{i\in T}(z_{i,\sigma(i)}z^*_{i,\sigma(i)})\prod_{i\in T^c}(z_{i,\sigma(i)}z^*_{i,\rho(i)})
    =\prod_{i\in T}h_1(z_{i,\sigma(i)})\prod_{i\in T^c}h_2(z_{i,\sigma(i)},z_{i,\rho(i)}).
\end{align}
Here, we assign the degree by adding 0 for $h_1$ and 1 for $h_2$.
Thus, the degree depends only on $T$, and we can rewrite the summation as
\begin{align}
    |\text{Per}(Z)|^2
    =\sum_{\sigma,\rho}\prod_{i=1}^N z_{i,\sigma(i)}z^*_{i,\rho(i)}
    =\sum_{k=0}^N \sum_{T\subset [N],|T_1|=|T_2|=k}\sum_{\sigma,\sigma',\rho'}\prod_{i\in T_1}h_1(z_{i,\sigma(i)})\prod_{i\in T_1^c}h_2(z_{i,\sigma'(i)},z_{i,\rho'(i)}),
\end{align}
where $\sigma \in \mathcal{S}_{k}$ is for $h_1$ and $\sigma', \rho'\in\mathcal{S}_{N-k}$ are for $h_2$, and $\sigma'(i)\neq \rho'(i)$ for all $i\in T_1^c$.
Also, here, $T_1$ and $T_2$ are the domain and range of $\sigma$, and $\sigma'$ and $\rho'$ are permutations from $T_1^c$ to $T_2^c$.
Thus, $\sigma\in\mathcal{S}_{|T_1|}$.
Thus, the $(N-k)$th order part is written as
\begin{align}
    g^{=(N-k)}
    &=\sum_{\substack{T,T'\subset [N] \\ |T|=|T'|=k}}\sum_{\substack{\sigma\in \mathcal{S}_k: \\ T\to T'}}\sum_{\substack{\sigma',\rho'\in \mathcal{S}_{N-k}: \\\sigma'(i)\neq \rho'(i), \\ T^c\to T'^{c}}}\prod_{i\in T}h_1(z_{i,\sigma(i)})\prod_{i\in T^c}h_2(z_{i,\sigma'(i)},z_{i,\rho'(i)}).
\end{align}
Using the orthogonality of the polynomials, we have
\begin{align}
    \mathbb{E}_Z[|g^{=(N-k)}|^2]
    =\binom{N}{k}^2(N-k)!(!(N-k))\sum_{j=0}^k\binom{k}{j}^2j!(k-j)!(!(k-j))2^j.
\end{align}
Here we used the fact that if two $T$'s are different, then the summands cannot be the same.
Thus, we can assume that the $T$'s from the left and right for the two-norm are the same.
Now, we have, for $|T|=k$,
\begin{align}
    \sum_{\sigma,\rho}\prod_{i\in T}h_1(z_{i,\sigma(i)})h_1(z_{i,\rho(i)})
    &=\sum_{\sigma,\rho}\prod_{i\in R}h_1(z_{i,\sigma(i)})^2\prod_{i\in T\setminus R}h_1(z_{i,\sigma(i)})h_1(z_{i,\rho(i)}) \\ 
    &=\sum_{R\subset T}\sum_{\sigma,\sigma',\rho'}\prod_{i\in R}h_1(z_{i,\sigma(i)})^2\prod_{i\in T\setminus R}h_1(z_{i,\sigma'(i)})h_1(z_i,\rho'(i)) \\ 
    &=\sum_{j=0}^k\binom{k}{j}^2j!(k-j)!(!(k-j))2^j,
\end{align}
where $\binom{k}{j}$ comes from counting $R$ and $R'$, which are the domain and range of $\sigma$.
And $2^j$ comes from $h_1^2$ and $j!$ for $\sigma$, $(k-j)!$ for $\sigma'$ and $!(k-j)$ for $\rho'$.
Here, $!k$ represents the number of derangements for $k$ elements, i.e., the number of permutations such that no elements appear in its original position.
Finally,
\begin{align}
    \sum_x \mathbb{E}_U[p_U(x)q_U(x)]
    &=\binom{M}{N}\frac{1}{M^{2N}} \sum_{k=0}^Nx^{N-k}\mathbb{E}_U[\|g^{=(N-k)}\|^2] \\ 
    &=\binom{M}{N}\frac{1}{M^{2N}} \sum_{k=0}^N x^{N-k}\binom{N}{k}^2(N-k)!(!(N-k))\sum_{j=0}^k\binom{k}{j}^2j!(k-j)!(!(k-j))2^j \\
    &\leq \binom{M}{N}\frac{1}{M^{2N}} \sum_{k=0}^N x^{N-k}e^2(N!)^2 \\ 
    &\leq \frac{e^2N!}{M^{N}}\frac{1-x^{N+1}}{1-x}.
\end{align}

\section{Analytic evidence supporting the numerical results}\label{app:evidence}
We consider FBS with $N$ photons and a linear-optical circuit $U$.
The probability of obtaining $N$ photons at the first $N$ output modes is written as
\begin{align}
    p(x)=|\per U_{N,N}|^2.
\end{align}
We now consider a mock-up distribution $h(x)$, which is the multinomial with uniform mixed input in Sec.~\ref{app:mock-up} as a 
\begin{align}
    h(x)=N!\prod_{i=1}^N\left(\frac{1}{N}\sum_{j=1}^N |U_{j,i}|^2\right).
\end{align}
Our goal is to compute XE between $p(x)$ and $h(x)^s$ with proper normalization:
\begin{align}
    \mathbb{E}_U[\text{XE}(p_U,h^s_U)]
    &\equiv \mathbb{E}_U\left[\frac{\sum_{x} p_U(x)h^s_U(x)}{\sum_{x} h^s_U(x)}\right].
\end{align}
We will approximate it by
\begin{align}
    \mathbb{E}_U[\text{XE}(p_U,h^s_U)]
    \approx \frac{\mathbb{E}_U[\sum_{x} p_U(x)h^s_U(x)]}{\mathbb{E}_U[\sum_{x} h^s_U(x)]},
\end{align}
which holds when the fluctuation of $\sum_x h^s_U(x)$ over $U$ is small and is exact when $s=1$.
One can expect that if $s$ is too large, the exponentiation with the power $s$ will significantly increase the fluctuation.
Thus, it only holds for small $s$.

Again, we will use the fact that for a sufficiently large number of modes $M=\omega(N^5)$, the $N\times N$ submatrix of a Haar-random unitary matrix $U\in\mathbb{C}^{M\times M}$ can be approximated by random Gaussian matrices whose elements follow the normal distribution with variance $1/M$ \cite{aaronson2011computational}.
More precisely, the element $U_{i,j}$'s real and imaginary parts follow the normal distribution with variance $1/(2M)$.
Defining random Gaussian matrices $Z\in \mathbb{C}^{N\times N}$ with variance $1/2$ and normalized variables, we can rewrite the XE as
\begin{align}
    \mathbb{E}_U[\text{XE}(p_U,h^s_U)]
    \approx \frac{\mathbb{E}_U[\sum_{x} p_U(x)h^s_U(x)]}{\mathbb{E}_U[\sum_{x} h^s_U(x)]}
    \approx \frac{N!\mathbb{E}_Z[\tilde{p}(Z)\tilde{h}^s(Z)]}{M^N\mathbb{E}_Z[ \tilde{h}^s(Z)]},
\end{align}
where
\begin{align}
    \tilde{p}(Z)=\frac{|\per Z|^2}{N!},~~~
    \tilde{h}(Z)=\frac{1}{N^N}\prod_{i=1}^N\sum_{j=1}^N |Z_{j,i}|^2,
\end{align}
which has a property $\mathbb{E}_Z[\tilde{p}(Z)]=\mathbb{E}_Z[\tilde{h}(Z)]=1$, where the average is taken over random Gaussian matrices $Z$.

%We will compute the XE between $\tilde{p}$ and $\tilde{h}^s$ for random Gaussian random matrices.

%From Ref.~\cite{aaronson2011computational}, we can also show that $\langle \tilde{p}(Z)^2\rangle=N+1$.
%Similarly, we can show that
%\begin{align}
%    \langle \tilde{p}(Z)\tilde{h}(Z)\rangle=\left(1+\frac{1}{N}\right)^N\approx e.
%\end{align}

We later show that
\begin{align} \label{eq:eqs}
    \mathbb{E}_Z[\tilde{h}^s(Z)]
    =\frac{1}{N^{sN}}\left(\frac{(N+s-1)!}{(N-1)!}\right)^N,~~~~~
    \mathbb{E}_Z[\tilde{p}(Z)\tilde{h}^s(Z)]
    =\frac{1}{N^{sN}}\left(\frac{(N+s)!}{N!}\right)^N.
\end{align}
Therefore, the XE of $\tilde{h}_U^s$ becomes
\begin{align}
    \mathbb{E}_U[\text{XE}(p_U,h_U^s)]
    \approx \frac{N!\mathbb{E}_Z[\tilde{p}(Z)\tilde{h}^s(Z)]}{M^N\mathbb{E}_Z[\tilde{h}(Z)^s]}
    =\frac{N!}{M^N}\left(1+\frac{s}{N}\right)^N\to e^s\frac{N!}{M^N}.
\end{align}
%Recall that
%\begin{align}
%    \text{XE}(\tilde{p},\tilde{p})
%    =\langle \tilde{p}(Z)^2\rangle
%    =N+1.
%\end{align}
%Thus, by choosing large $s$, we can achieve large XE.
%We emphasize that we are computing XE between two functions with respect to {\it random Gaussian matrices}, which might be different from the XE of the probability distributions.
We emphasize that although we considered $h^s$, we do not have an efficient method of sampling the corresponding distribution $h^s$.
The point of employing this distribution is to effectively simulate the post-selection of heavy outcomes by taking the $s$th power.

\iffalse
We want to compute
\begin{align}
    \left\langle \sum_{x} p_U(x)\frac{h^s_U(x)}{\sum_{x} h^s_U(x)}\right\rangle_U
    =\left\langle \frac{\sum_{x} p_U(x) h^s_U(x)}{\sum_{x} h^s_U(x)}\right\rangle_U
    \equiv \left\langle\frac{f(U)}{g(U)}\right\rangle_U
    \approx \frac{\bar{f}}{\bar{g}}-\frac{\text{Cov}(f,g)}{\bar{g}^2}+\frac{\bar{f}\text{Var}(g)}{\bar{g}^3},
\end{align}
where the expansion is for small $\text{Var}(g)$ and $\text{Cov}(f,g)=\langle fg\rangle_U-\langle f\rangle_U\langle g\rangle_U$.

\begin{align}
    \text{Cov}(f,g)
    =\left\langle \sum_x p_U(x)h_U^s(x) \sum_y h_U^s(y) \right\rangle_U
    -\bar{f}\bar{g}.
\end{align}

\begin{align}
    \text{Var}(g)
    =\left\langle \sum_x h_U^s(x) \sum_y h_U^s(y) \right\rangle_U
    -\bar{g}^2.
\end{align}
\fi

\begin{figure*}[t]
\includegraphics[width=500px]{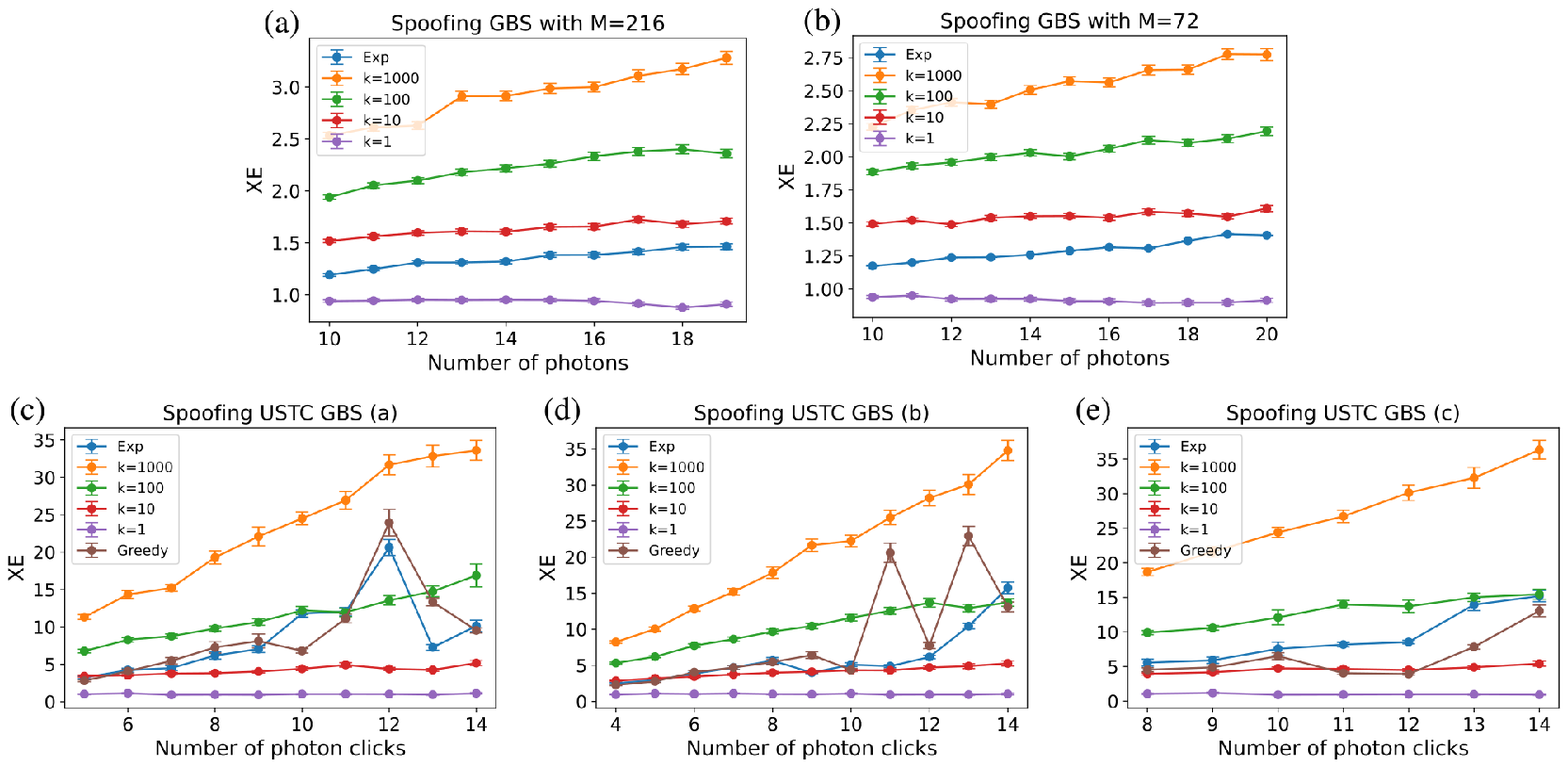}
\caption{Spoofing linear XE test instead of log XE for the experiments in Refs.~\cite{zhong2021phase, madsen2022quantum}}.
\label{fig:linear}
\end{figure*}

\begin{widetext}
Now we derive Eq.~\eqref{eq:eqs}.
From now on, we will drop the tilde of $\tilde{h}$ and $\tilde{p}$ for simplicity.
Let us first compute
\begin{align}
    \mathbb{E}_Z[h(Z)^s]
    %=\frac{1}{(N^N)^s}\left\langle \left[\prod_{j=1}^N\left(\sum_{i_j=1}^N|Z_{j,i_j}|^2\right)\right]^s\right\rangle
    =\frac{1}{N^{sN}}\mathbb{E}_Z\left[\prod_{j=1}^N\left(\sum_{i_j=1}^N|Z_{j,i_j}|^2\right)^s\right]
    =\frac{1}{N^{sN}}\mathbb{E}_Z\left[\left(\sum_{i_j=1}^N|Z_{j,i_j}|^2\right)^s\right]^N.
\end{align}
Here,
\begin{align}
    \frac{1}{N^{sN}}\mathbb{E}_Z\left[\left(\sum_{i_j=1}^N|Z_{j,i_j}|^2\right)^s\right]^N
    &=\frac{1}{N^{sN}}\left(\sum_{\bm{k}:\sum_i k_i=s}\frac{s!}{k_1!\cdots k_N!}\mathbb{E}_z [z^{2k_1}]\cdots \mathbb{E}_z[z^{2k_N}]\right)^N \\ 
    &=\frac{1}{N^{sN}}\left(s!\sum_{\bm{k}:\sum_i k_i=s}\right)^N
    =\frac{1}{N^{sN}}\left(s!\binom{N+s-1}{N-1}\right)^N
    =\frac{1}{N^{sN}}\left(\frac{(N+s-1)!}{(N-1)!}\right)^N,
\end{align}
where we have used for $z=x+iy$
\begin{align}
    \mathbb{E}_z[z^{2s}]
    &=\mathbb{E}_z[(x^2+y^2)^s]
    =\sum_{l=0}^s\binom{s}{l}\mathbb{E}_z [x^{2l}]\mathbb{E}_z[y^{2(s-l)}]
    =\sum_{l=0}^s\binom{s}{l}\mathbb{E}_z [x^{2}]^s(2l-1)!!(2(s-l)-1)!! \\ 
    &=\sum_{l=0}^s\binom{s}{l}\frac{1}{2^s}\frac{(2l)!}{2^ll!}\frac{(2(s-l))!}{2^{s-l}(s-l)!} 
    =\sum_{l=0}^s\binom{s}{l}\frac{1}{4^s}\frac{(2l)!}{l!}\frac{(2(s-l))!}{(s-l)!}
    =s!,
\end{align}
and
\begin{align}
    \sum_{\bm{k}:\sum_k k_i=s}=\binom{N+s-1}{N-1}.
\end{align}

Now, let us compute
\begin{align}
    \mathbb{E}_Z[p(Z)h(Z)^s]
    &=\frac{1}{(N^N)^s}\mathbb{E}_Z\left[
    \frac{1}{N!}\sum_\sigma \prod_{j=1}^N|Z_{j,\sigma(j)}|^2\left[\prod_{j=1}^N\left(\sum_{i_j=1}^N|Z_{j,i_j}|^2\right)\right]^s\right]
    =\frac{1}{N^{sN}}\mathbb{E}_Z\left[\prod_{j=1}^N \left[|Z_{j,j}|^2\left(\sum_{i_j=1}^N|Z_{j,i_j}|^2\right)^s\right]\right] \\ 
    &=\frac{1}{N^{sN}}\mathbb{E}_Z\left[ |Z_{1,1}|^2\left(\sum_{i=1}^N|Z_{1,i}|^2\right)^s\right]^N \label{eq:power_s},
\end{align}
where the last equality is from the independence of elements in random matrix $Z$.
Now, we count the number of $i$'s such that $i=1$ after expanding the $s$th power.
First, when there is no $i_j$ for all $j\in [s]$, then we have
\begin{align}
    \mathbb{E}_Z[|Z_{1,1}|^2]\mathbb{E}_Z\left[\left(\sum_{i=2}^{N}|Z_{1,i}|^2\right)^s\right]
    =s!\binom{N+s-2}{N-2}.
\end{align}
When there is one $i$ such that $i=1$, we have
\begin{align}
    \binom{s}{1}\mathbb{E}_Z[|Z_{1,1}|^4]\mathbb{E}_Z\left[\left(\sum_{i=2}^{N}|Z_{1,i}|^2\right)^{s-1}\right]
    =\binom{s}{1}2!(s-1)!\binom{N+s-3}{N-2}.
\end{align}
When there are $l$ $i$'s such that $i=1$, we have
\begin{align}
    \binom{s}{l}\mathbb{E}_Z[|Z_{1,1}|^{2(l+1)}]\mathbb{E}_Z\left[\left(\sum_{i=2}^{N}|Z_{1,i}|^2\right)^{s-l}\right]
    =\binom{s}{l}(l+1)!(s-l)!\binom{N+s-l-2}{N-2}.
\end{align}
Thus, we can finally simplify the expression in Eq.~\eqref{eq:power_s}
\begin{align}
    \mathbb{E}_Z\left[|Z_{1,1}|^2\left(\sum_{i=1}^N|Z_{1,i}|^2\right)^s\right]
    =\sum_{l=0}^s \binom{s}{l}(l+1)!(s-l)!\binom{N+s-l-2}{N-2}
    =\frac{(N+s)!}{N!}
\end{align}
Hence,
\begin{align}
    \mathbb{E}_Z[p(Z)h(Z)^s]
    =\frac{1}{N^{sN}}\left(\frac{(N+s)!}{N!}\right)^N.
\end{align}

\section{Linear cross entropy}\label{app:linear}
We now present the result in the main text by replacing log XE with linear XE:
\begin{align}
    \text{XE}=\sum_xq(x)p(x).
\end{align}
Note that $\log p(x)$ is replaced by $p(x)$.
The results are exhibited in Fig.~\ref{fig:linear}.
From the figure, one can see that the trend does not differ between linear XE and log XE.

\iffalse
For a constant $s$, we have
\begin{align}
    \left(1+\frac{s}{N}\right)^N\to e^s,
\end{align}
and if we choose $s=N$, we have 
\begin{align}
    \left(1+\frac{s}{N}\right)^N= 2^N,
\end{align}
and if $s\gg N$,
\begin{align}
    \left(1+\frac{s}{N}\right)^N\approx \left(\frac{s}{N}\right)^N.
\end{align}
\fi

\end{widetext}

%\clearpage

\bibliography{reference.bib}